\documentclass[10pt,twocolumn,letterpaper]{article}
\usepackage[accsupp]{axessibility}  %

\usepackage{cvpr}              %

\usepackage[dvipsnames]{xcolor}

\definecolor{cvprblue}{rgb}{0.21,0.49,0.74}
\usepackage[pagebackref,breaklinks,colorlinks,citecolor=cvprblue]{hyperref}
\def\paperID{1505} %
\def\confName{CVPR}
\def\confYear{2025}

\usepackage[utf8]{inputenc} %
\usepackage[T1]{fontenc}    %
\usepackage{hyperref}       %
\usepackage{url}            %
\usepackage{booktabs}       %
\usepackage{amsfonts}       %
\usepackage{nicefrac}       %
\usepackage{microtype}      %
\usepackage{xcolor}         %
\usepackage{booktabs} 
\usepackage{amsmath}
\usepackage{float}          %
\usepackage{graphicx}
\graphicspath{{media}}
\usepackage{subcaption}
\usepackage{enumitem}
\usepackage{multirow}
\usepackage{amssymb}%
\usepackage{pifont}%
\newcommand{\cmark}{\ding{51}}%
\newcommand{\xmark}{\ding{55}}%

\newcommand{\x}{\mathbf{x}}

\newcommand{\K}{\mathbf{k}}
\newcommand{\ft}{\mathcal{F}}

\newcommand{\ino}{$\text{NO}_{\textbf{i}}$\xspace}
\newcommand{\kno}{$\text{NO}_{\textbf{k}}$\xspace}

\newcommand{\norm}[1]{\left\lVert#1\right\rVert}

\DeclareMathOperator{\argmin}{argmin}
\DeclareMathOperator{\ssim}{SSIM}
\DeclareMathOperator{\CNN}{CNN}

\usepackage{amsthm}
\theoremstyle{plain}
\newtheorem{theorem}{Theorem}[section]

\theoremstyle{definition}
\newtheorem{definition}[theorem]{Definition}
\theoremstyle{remark}

\title{A Unified Model for Compressed Sensing MRI Across Undersampling Patterns}
\author{%
  Armeet Singh Jatyani$^*$ \quad\quad Jiayun Wang$\thanks{Equal contribution.}$  \quad\quad Aditi Chandrashekar \quad\quad Zihui Wu \\
  {Miguel Liu-Schiaffini} \quad\quad\quad\; {Bahareh Tolooshams} \quad \quad \quad\; {Anima Anandkumar} \\
  California Institute of Technology \\
\small  \texttt{\{armeet,peterw,ajchandr,zwu2,mliuschi,btoloosh,anima\}@caltech.edu} \\
}

\begin{document}
\maketitle

\begin{abstract}
Compressed Sensing MRI reconstructs images of the body's internal anatomy from undersampled measurements, thereby reducing scan time—the time subjects need to remain still.
Recently, deep learning has shown great potential for reconstructing high-fidelity images from highly undersampled measurements. However, one needs to train multiple models for different undersampling patterns and desired output image resolutions, since most networks operate on a fixed discretization.  
Such approaches are highly impractical in clinical settings, where undersampling patterns and image resolutions are frequently changed to accommodate different real-time imaging and diagnostic requirements.

We propose a unified MRI reconstruction model robust to various measurement undersampling patterns and image resolutions.  Our approach uses neural operators—a discretization-agnostic architecture applied in both image and measurement spaces—to capture local and global features. Empirically, our model improves SSIM by 11\% and PSNR by $4$ dB over a state-of-the-art CNN (End-to-End VarNet), with 600$\times$ faster inference than diffusion methods. The resolution-agnostic design also enables zero-shot super-resolution and extended field-of-view reconstruction, offering a versatile and efficient solution for clinical MR imaging. 
Our unified model offers a versatile solution for MRI, adapting seamlessly to various measurement undersampling and imaging resolutions, making it highly effective for flexible and reliable clinical imaging. Our code is available at \url{https://armeet.ca/nomri}.

\end{abstract}

\begin{figure*}[t!]
    \centering
    \includegraphics[width=\textwidth]{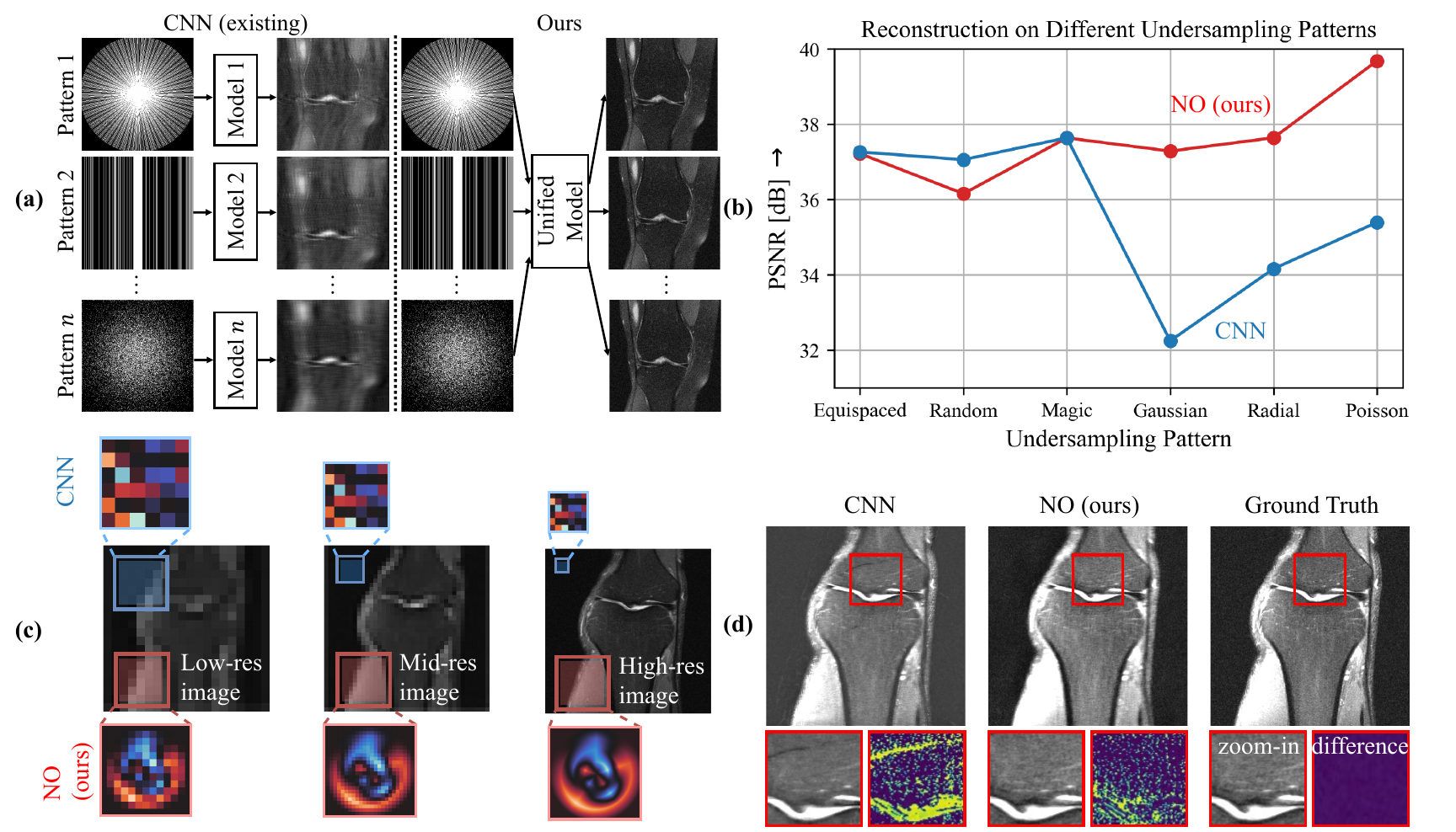}
    \caption{  
    {\bf (a) We propose a unified model for MRI reconstruction, called neural operator (NO)}, which works across various measurement undersampling patterns, overcoming the resolution dependency limit of CNN-based methods like \cite{e2evarnet} that require a specific model for each pattern.
{\bf (b) NO achieves consistent performance across undersampling patterns} and outperforms CNN architectures such as \cite{e2evarnet} (for 2$\times$ acceleration with one unrolled network cascade).
{\bf (c) NO is resolution-agnostic}. As image resolution increases, it maintains a consistent kernel size for alias-free rescaling, unlike CNNs with variable kernel sizes that risk aliasing.
    {\bf (d) NO enhances zero-shot super-resolution} MRI reconstruction, outperforming CNNs \cite{e2evarnet}.}
    \label{fig:intro}
    \vspace{-1em}
\end{figure*}

\section{Introduction}
\label{sec:intro}

Magnetic Resonance Imaging (MRI) is a popular non-invasive imaging technology, used in numerous medical and scientific applications such as neurosurgery~\cite{seifert1999open}, clinical oncology~\cite{koh2007diffusion}, diagnostic testing \cite{husband2001mri}, neuroscience~\cite{le2003looking}, and pharmaceutical research~\cite{richardson2005pharmaceutical}. MRI is greatly limited by a slow data acquisition process, which sometimes requires patients to remain still for an hour \cite{chen2022ai,singh2023emerging}. Hence, accelerating MRI scan has garnered tremendous attention \cite{griswold2002generalized,lustig2008compressed,johnson2019conditional}. %

Compressed Sensing (CS) \cite{donoho2006cs} enables MRI at sub-Nyquist rates and reduces acquisition time for greater clinical utility. This is framed as an ill-posed inverse problem \cite{groetsch1993inverse}, where prior knowledge about MR images is crucial for reconstruction. Traditional Compressed Sensing MRI %
assumes a sparse prior in a transform domain (e.g., wavelets \cite{chen2001atomic}). Recent deep learning methods learn underlying data structures to achieve superior performance~\cite{e2evarnet,score_sde}. Current state-of-the-art models establish an \emph{end-to-end} mapping \cite{e2evarnet,varnet4} from undersampled measurements to image reconstruction in both image and frequency domains. However, these models often struggle with generalization across varying resolutions, a critical need in clinical practice where flexible resolution adjustments are necessary. A unified model that is agnostic to discretizations would greatly improve efficiency. %

Neural Operators (NOs) \cite{nop} are a deep learning framework that learns mappings between infinite-dimensional function spaces, making them agnostic to discretizations (resolutions). This property makes them suitable for tasks with data at varying resolutions, such as partial differential equations (PDEs) \cite{nop,li2024physics,raonic2024convolutional} and PDE-related applications \cite{rashid2022learning,pathak2022fourcastnet}. NOs could also be suitable for compressed sensing MRI due to measurements with multiple undersampling patterns. 
Various NO architectures \cite{fnop,raonic2024convolutional,local_no} have been proposed. Recently, discrete-continuous (DISCO) convolutions  \cite{local_no,ocampo2022} have emerged as an efficient neural operator that captures local features and leverages GPU acceleration for standard convolutions.  
Due to the similarity to standard convolutions, the building blocks of many existing MRI deep learning models \cite{e2evarnet,score_sde}, DISCO is a good candidate for resolution-agnostic MRI reconstruction.

{\bf Our approach:}
 We propose a unified model based on NOs, that is robust to different undersampling patterns and image resolutions in compressed sensing MRI (Fig.~\ref{fig:intro}a).
 Our model follows an unrolled network design \cite{varnet4,e2evarnet} with DISCO  \citep{ocampo2022,local_no}. %
As the image resolution increases,  DISCO maintains a resolution-agnostic kernel with a consistent convolution patch size, while the regular convolution kernel contracts to a point (Fig.~\ref{fig:intro}c). The DISCO operators learn in both measurement/frequency $\K$ space (\kno) and image space (\ino).
\kno makes our framework agnostic to different measurement undersampling patterns, and \ino makes the framework agnostic to different image resolutions. Additionally, the learning in both frequency and image space allows the model to capture both local and global features of images due to the duality of the Fourier transform that connects the frequency and image space. %
 The resolution-agnostic design also enables super-resolution in both frequency and image space, allowing the extended field of view (FOV) and super-resolution of the reconstructed MR images.  

We empirically demonstrate that our model is robust to different measurement undersampling rates and patterns (Fig.~\ref{fig:intro}a). Our model performs consistently across these pattern variations, whereas the existing method drops in performance (Fig.~\ref{fig:intro}b). We achieve up to 4$\times$ lower NMSE and 5 dB PSNR improvement from the baseline when evaluating on different undersampling patterns. The model is efficient and $600 \times$ faster than the diffusion baseline \cite{jalal2021robust,wu2024principled,score_sde}. We also show that our model outperforms the state-of-the-art in zero-shot super-resolution inference (Fig.~\ref{fig:intro}d) and extended FOV reconstruction (Fig.~\ref{fig:superres}).
 
Our work has two main contributions: {\bf 1)} We propose a unified neural operator model that learns in function space and shows robust performance across different undersampling patterns and image resolutions in compressed sensing MRI. To the best of our knowledge, this is the first resolution-agnostic framework for MRI reconstruction. %
{\bf 2)} Our model demonstrates empirical robustness across measurement undersampling rates and patterns, reconstructing MR images with zero-shot higher resolutions and a larger field of view.%

\section{Related Works}
\label{sec:related}

{\bf Accelerated MRI}.
One way to accelerate MRI scan speed is parallel imaging, in which multiple receiver coils acquire different views of the object of interest simultaneously, and then combine them into a single image \cite{griswold2002generalized,murphy2012fast,ronneberger2015u}. 
When MRI reconstruction is paired with compressed sensing, pre-defined priors or regularization filters can be leveraged to improve reconstruction quality \cite{lustig2008compressed,cs_mri}.
Recent works have shown that learned deep-learning priors outperform hand-crafted priors in reconstruction fidelity.  
Convolutional neural networks (CNNs)~\cite{johnson2019conditional, varnet4, darestani2021accelerated, e2evarnet}, variational networks (based on variational minimization)~\cite{varnet4, e2evarnet}, and generative adversarial networks (GANs)~\cite{johnson2019conditional, dar2020prior} have all demonstrated superior performance than traditional optimization approach for compressed sensing MRI reconstruction from undersampled measurements.  
However, unlike conventional compressed sensing which operates in the function space and is agnostic to measurement undersampling patterns, the aforementioned deep learning methods operate on a fixed resolution.
As a result, changes in resolution lead to degradation in performance, and multiple models are needed for different settings. We propose a resolution-agnostic unified model.

\noindent{\bf Discretization-Agnostic Learning and Neural Operators}.
Empirically,  diffusion models have shown relatively consistent performance with different measurement undersampling patterns in accelerated MRI  \cite{gungor2023adaptive}. However, diffusion models usually take more runtime at inference and need extensive hyperparameter tuning for good performance (Section \ref{sec:analysis}). Additionally, they are not fundamentally discretization-agnostic by design. 
Neural operators \cite{azizzadenesheli2024neural,nop} are deep learning architectures specifically designed to learn mappings between infinite-dimensional function spaces. They are discretization-agnostic, allowing evaluation at any resolution, and converge to a desired operator as the resolution approaches infinity.
Neural operators have empirically achieved good performance as surrogate models of numerical solutions to partial differential equations (PDEs) \cite{nop,li2024physics,raonic2024convolutional} with various applications, such as material science~\cite{rashid2022learning}, weather forecasting~\cite{pathak2022fourcastnet}, and photoacoustic imaging~\cite{guan2023fourier}. %
The design of neural operators often depends on the application at hand. 
For example, the Fourier neural operator (FNO) \cite{fnop}, which performs global convolutions, has shown consistent discretization-agnostic performance in various applications \cite{azizzadenesheli2024neural}.
Other designs of neural operators \cite{gnop,local_no} rely on integration with locally-supported kernels to capture local features, which has shown to be useful in applications where local features are important, such as modeling turbulent fluids \cite{gnop}. Additionally, neural operators with local integrals can be made efficient with parallel computing compared to those requiring global integrals.
Our MRI framework, based on neural operators with local integrals, is agnostic to undersampling patterns and output image resolutions.

\begin{figure*}[t!]
\vspace{-1em}
    \centering
    \includegraphics[width=1\linewidth]{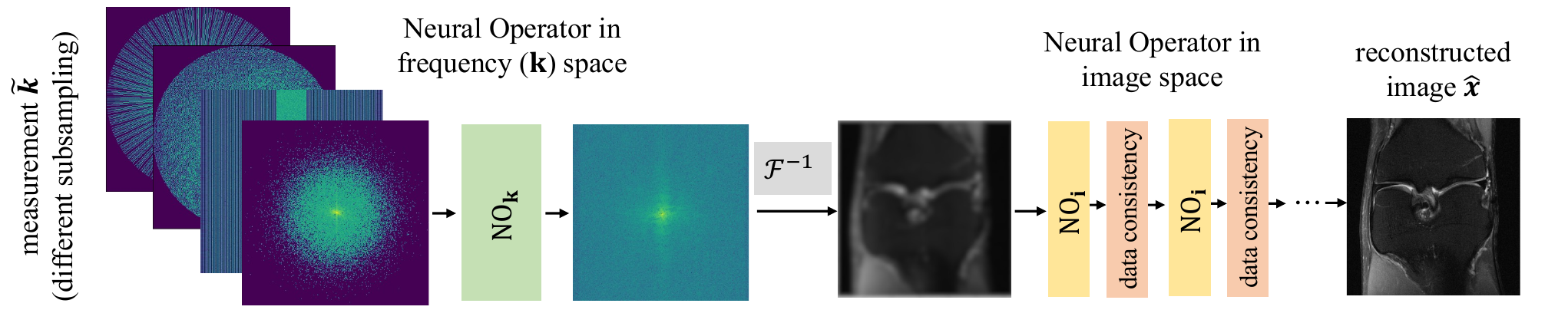}
    \caption{ {\bf MRI reconstruction pipeline.} NO learns data priors in function space with infinite resolution. Specifically we propose 
   NOs in the $\K$ (frequency) space \kno ($\K$ space NO) and image space \ino (image space NO), which capture both global and local image features, due to the duality between physical and frequency space. $\mathcal{F}^{-1}$ refers to the inverse Fourier transform. We provide the framework design details in Section \ref{sec:unrolled} and   NO design details in Section \ref{sec:nodesign}.
    }
    \vspace{-1em}
    \label{fig:framework}
\end{figure*}

\section{Methods}
We first discuss the background of compressed sensing MRI and the unrolled network framework we use. We then discuss how we can extend the existing network building block, standard convolution, to resolution-agnostic neural operators. %
We also introduce DISCO \citep{ocampo2022}, a neural operator design we adopt, and we capture global and local image features with DISCO. We conclude the section with the super-resolution designs. We call the measurement or frequency space \textit{$\K$-space}, and physical or spatial space \textit{image space} hereafter.
    
\subsection{MRI Reconstruction with Unrolled Networks}
\label{sec:unrolled}

\noindent {\bf Background}. In MRI, anatomical images $\x$ of the patient are reconstructed by acquiring frequency-domain measurements $\K$, where the relationship is defined as:
\begin{align}
    \K := \ft(\x) + \epsilon
\end{align}
where $\epsilon$ is the measurement noise and $\ft$ is the Fourier transform. 
In this paper, we consider the parallel imaging setting with multiple receiver coils \cite{juchem2015multi,zbontar2018fastMRI} , where each coil captures a different region of the anatomy. 
The forward process of the $i^{\text{th}}$ coil measures $\K_i:=\ft(S_i \x)+ \epsilon_i$ where $S_i$ is a position-dependent sensitivity map for the $i^{\text{th}}$ coil.%
~To speed up the imaging process, measurements are undersampled as $\tilde{\K}=M\K$ in the compressed sensing MRI setting, where $M$ is a binary mask that selects a subset of the k-space points. 
Classical compressed sensing methods reconstruct the image $\hat{\x}$ by solving an optimization problem  
\begin{align}\label{eq:opt}
    \hat{\x} = \argmin_{\x} \frac{1}{2} \sum_{i} \norm{ \mathcal{A}( \x) - \tilde{\K}}_2^2 + \lambda \Psi(\x)
\end{align}
where $i$ is the coil index, $\mathcal{A}(\cdot) := M \mathcal{F}S_i(\cdot)$ is the linear forward operator, and $\Psi(\x)$ is a regularization term. The optimization objective can be considered as a combination of physics constraint and prior.
While the above optimization can be solved using classical optimization toolboxes, an increasing line of works uses deep neural networks to learn data priors and show improved reconstruction performance \cite{varnet4,e2evarnet}. 
Among them, unrolled networks~\cite{varnet4,e2evarnet} have gained popularity as they incorporate the known forward model, resulting in state-of-the-art performance. Unrolling, which started with the nominal work of LISTA~\cite{gregor2010learning}, proposes to design networks using iterations of an optimization algorithm to solve inverse problems. 
This approach incorporates domain knowledge (i.e., the forward model) and leverages deep learning to learn implicit priors from data~\cite{sun2016deep, mardani2018neural}.
In the context of MRI and assuming a differential regularization term, the optimization problem is expanded to iterative gradient descent steps with injected CNN-based data priors.
Each layer mimics the gradient descent step from $\x^{t}$ to $\x^{t+1}$:
\begin{align}
    \x^{t+1} &\leftarrow  \x^t-\eta^t\mathcal{A}^*(\mathcal{A}(\x^t)-\tilde{\K})+ \lambda^t \CNN(\x^t)%
\end{align}
where $\eta^t$ controls the weight of {\it data consistency} term and $\lambda^t$ controls that of the data-driven prior term. %
The data consistency term samples the data in the frequency domain, hence it is applicable to any spatial resolution. However, the prior term only operates on a specific resolution with CNNs. This means when changing the undersampling patterns, one needs another CNN trained for that setting, which greatly limits the flexibility of the reconstruction system. 

\noindent{\bf Extending to Neural Operators}. We learn the prior in function space via discretization-agnostic neural operators in $\K$ space (\kno) and image space (\ino). 
Specifically, we first use a $\K$ space neural operator \kno to learn $\K$ space prior and then apply a cascade of  unrolled layers, each of which features a data consistency loss and the image space \ino for image prior learning:  
\begin{align}
\label{eq:no}
        \x^0 &\leftarrow \mathcal{F}^{-1}(\text{\kno}(\tilde{\K})) \\
\label{eq:no2}
        \x^{t+1} &\leftarrow \x^t-\eta^t\mathcal{A}^*(\mathcal{A}(\x^t)-\tilde{\K})+ \lambda^t \text{NO}_{\textbf{i}}^t (\x^t) %
\end{align}
where $\text{NO}_{\textbf{i}}^t$ refers to the image-space NO at cascade $t$. 
We follow existing works \cite{varnet4,e2evarnet} and only have one \kno for the first cascade. %
Our framework flexibly works for different resolutions with the design details in Section \ref{sec:nodesign}.%

\noindent\textbf{Framework Overview}. Fig.~\ref{fig:framework} depicts the pipeline of our neural operator framework for MRI reconstruction.
The undersampled measurement $\tilde{\K}$ is first fed to a neural operator \kno which operates in measurement $\K$ space to learn global image features and then inverse Fourier transformed to get an image.
Following Eqn.~\ref{eq:no} and \ref{eq:no2}, we iterate a few cascades of unrolled layers, consisting of a neural operator \ino which operates in image $\x$ space and a data consistency update.

\subsection{Neural Operator Design}
\label{sec:nodesign}
Neural operators, which learn mappings between function spaces, offer a unified approach to discretization-agnostic MRI reconstruction. Given that accurate MRI reconstruction depends on capturing both local and global image features, we propose a neural operator architecture that incorporates both global and local inductive biases. 
We first discuss how we learn local features with local integration operators.

\noindent {\bf Local Features via Local Integration Operator}. %
Historically, the most common method of embedding a local inductive bias into deep neural networks has been by using locally-supported convolutional kernels, as in convolutional neural networks (CNNs). 
However, standard discrete convolutional kernels used in CNNs do not satisfy the resolution-agnostic properties of neural operators. 
Specifically, Liu et al. \cite{local_no} show that CNN-style convolutional kernels converge to pointwise linear operators as the resolution is increased, instead of the desired local integration in the limit of infinite resolution. 
For a kernel $\kappa$ and input function $g$ defined over some compact subset $D \subset \mathbb{R}^d$, the \textit{local convolution operator} in a standard convolution layer, which transforms input $u$ to output $v$, is given by
\begin{equation}
\label{eq:conv}
    (k \star g)(v) = \int_{D} \kappa(u - v) \cdot g(u) \ \mathrm{d}u.
\end{equation}
Given a particular set of input points $(u_j)_{j=1}^m \subset D$ with corresponding quadrature weights $q_j$ and output positions $v_i \in D$, we adopt the discrete-continuous convolutions (DISCO) framework for operator learning~\citep{ocampo2022, local_no} and approximate the continuous convolution (Eqn. \ref{eq:conv}) as
\begin{equation}
    \label{eq:disc_conv}
    (k \star g)(v_i) \approx \sum_{j=1}^m \kappa(u_j - v_i) \cdot g(x_j) q_j.
\end{equation}
We follow parameterize $\kappa$ as a linear combination of pre-defined basis functions $\kappa^{\ell}$: $\kappa = \sum_{\ell = 1}^L \theta^{\ell} \cdot \kappa^{\ell}$, where $\theta^{\ell}$ are learnable parameters. We choose the linear piecewise basis from \cite{local_no} as this achieves the greatest empirical results (see Sections \ref{sec:udno_disco_implementation} \& \ref{sec:discodetails} of the supplementary). The convolutional kernel is thus parameterized by a finite number of parameters, independently of the grid on which the kernel is evaluated. The kernel is resolution-agnostic because we {\it disentangle} the resolution-agnostic basis and discrete learnable parameters. The basis $\kappa^{\ell}$ is defined in the function space, and will be discretized at the desired resolution; discrete parameters $\theta^{\ell}$ can be learned with gradient descent. 
Since we are operating on an equidistant grid on a compact subset of $\mathbb{R}^2$, we follow \cite{local_no} and 
implement Eqn.~\ref{eq:disc_conv} using standard convolutional kernels (thus enjoying the benefits of acceleration on GPUs using standard deep learning libraries) with two crucial modifications: {\bf 1)} the kernel itself is defined as a linear combination of basis functions $\kappa^{\ell}$, and {\bf 2)} the size of the kernel scales with the input resolution so as to remain a fixed size w.r.t. the input domain. We adopt the same basis functions as \cite{local_no} in our experiments, and we use the local integration operator as the resolution-agnostic building block for the measurement space and image space operators. 

\noindent {\bf DISCO vs Standard 2D Convolution with Varying Resolutions}. 
As the input resolution increases (the discretization becomes denser), DISCO \citep{ocampo2022} maintains the kernel size for each convolution and finally converges to a local integral. The standard 2D convolution kernel, however, gets increasingly smaller and finally converges to a point-wise operator (Fig.~\ref{fig:intro}c).
Although one could alleviate the issue of standard convolutions by interpolating the convolutional kernel shape to match with corresponding convolution patch sizes for different resolutions, the interpolated kernel will have artifacts that affect performance at new resolutions (Fig.~\ref{fig:intro}d). %
DISCO, however, is agnostic to resolution changes as the kernel is in the function space. 

\noindent {\bf Global Features}.
A common neural operator architecture for learning global features is the Fourier neural operator (FNO)~\cite{fnop}. FNO takes the Fourier transform of the input, truncates the result beyond some fixed number of modes, and pointwise multiplies the result with a learned weight tensor, which is equivalent to a global convolution on the input by the convolution theorem.
Interestingly, the forward process of MRI is a Fourier transformation, which means that local operations in measurement $\K$ space are equivalent to global operators in image $\x$ space and vice versa, due to their duality.
Following FNO, we could apply a pointwise multiplication between the measurement $\K$ and a learned weight tensor to capture global image features.  
However, FNO truncates high frequencies, which are crucial for MRI reconstruction.  
To address this, we directly apply the DISCO local integration operator on the measurement space to capture global image features without feature map truncation.

\noindent {\bf UDNO: the Building Block}. Without loss of generality, we make both the image-space \ino and $\K$ space \kno be local neural operators that capture local features in the corresponding domain. Such a design learns both global and local image features due to domain duality. Motivations for adopting the U-shaped architecture are in Fig.~\ref{fig:udnoa}  and Section \ref{sec:udno_a} of the Supplementary.
Each operator consists of multiple sub-layers, to which we refer as the U-Shaped DISCO Neural Operator, or UDNO. The motivation is that multi-scale designs have shown great success in capturing features at different scales in images and that U-shaped networks are among the most popular architectures in computer vision, demonstrating strong performance in various applications from medical imaging to diffusion \cite{ronneberger2015u,peebles2023scalable,croitoru2023diffusion}. Further, UDNO makes our framework very similar to an existing state-of-the-art E2E-VN \cite{e2evarnet}, with the difference being standard convolutions replaced by DISCO operators.  
The UDNO follows the encoder/decoder architecture of the U-Net \cite{ronneberger2015u}, replacing regular convolutions with DISCO layers.

\noindent {\bf Loss}. The parameters of the proposed neural operator are estimated from the training data by minimizing the structural similarity loss between the reconstruction $\x$ and the ground truth image $\x^*$ (the same as the E2E-VN \cite{e2evarnet}):
\vspace{-0.5em}
\begin{equation}
     \mathcal{L}(\hat{\x}, \x^*) = - \ssim(\hat{\x}, \x^*),
\end{equation}
where SSIM is the Structural Similarity Index Measure~\cite{wang2003multiscale}.%

\begin{figure}[t!]
    \centering
    \includegraphics[width=\columnwidth]{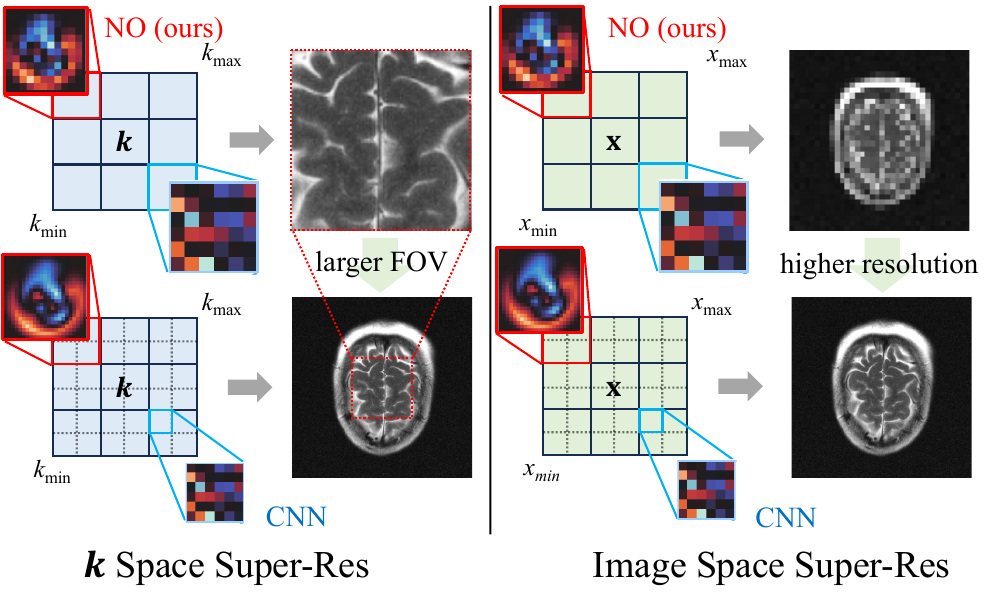}
    \vspace{-1em}
    \caption{{\bf Super resolution} (denser discretization) in $\K$ space or image space increases the FOV or resolution of the reconstructed image. With denser discretization,  NO maintains a resolution-agnostic kernel while CNN kernels become relatively smaller in size. Empirically our NO outperforms CNNs \cite{e2evarnet} (Section \ref{sec:superres}). }
    \label{fig:super_res_method}
\end{figure}

\subsection{Super-Resolution}

Neural operators enable zero-shot super-resolution. As shown in Fig.~\ref{fig:super_res_method}, increasing resolution corresponds to denser discretization between fixed minimum and maximum values, while the overall domain range remains constant. Due to the dual nature of frequency and image space, enhancing resolution in $\K$ space extends the field of view (FOV) in the reconstructed image, whereas increasing resolution in image space enhances the image’s detail. Our proposed NO framework includes resolution-agnostic neural operators for both $\K$ space (\kno) and image space (\ino), facilitating zero-shot super-resolution in both domains. We present empirical zero-shot super-resolution results in Section \ref{sec:superres}, comparing our NO framework to E2E-VN \cite{e2evarnet}, a CNN-based architecture with a similar design.

\section{Experiments}
We discuss the datasets and experimental setup, followed by comparisons of our and baseline methods with different $\K$ undersampling rates and patterns. We conclude the section with zero-shot super-resolution and additional analysis.
\label{sec:exp}

\subsection{Dataset and Setup}
\noindent{\bf Datasets}:
The fastMRI dataset \cite{zbontar2018fastMRI} is a large and open dataset of knee and brain fully-sampled MRIs. 
\begin{itemize}
    \item \textbf{fastMRI knee}:  We use the multi-coil knee reconstruction dataset with 34,742 slices for training and 7,135 slices for evaluation. All samples contain data from 15 coils. 
    \item \textbf{fastMRI brain}: We use the T2 contrast subset of the multi-coil brain reconstruction dataset with 6,262 training slices and 502 evaluation slices. We filter for samples with data from 16 coils.

\end{itemize}

\noindent{\bf Undersampling Patterns and Rates}. We use equispaced, random, magic, Gaussian, radial, and Poisson undersampling patterns~\cite{zbontar2018fastMRI, score_sde} and 2x, 4$\times$, 6$\times$, and 8$\times$ undersampling rates (visualizations are in Fig.~\ref{fig:mask_fns} in the Supplementary). Higher rates result in sparser $\K$ space samples and shorter imaging time at the cost of a more ill-posed/harder inversion process. Section \ref{sec:secb} in the Supplementary provides additional undersampling details along with mask visualizations.

\begin{table}[t]
    \centering
    \label{table:results_rate}
    \resizebox{\linewidth}{!}{
            \begin{tabular}{llcccc}
            \toprule
            \multirow{2}{*}{\textbf{Category}} &\multirow{2}{*}{\textbf{Method}} & \multicolumn{2}{c}{\textbf{fastMRI knee}} & \multicolumn{2}{c}{\textbf{fastMRI brain}} \\
            \cmidrule(lr){3-4} \cmidrule(lr){5-6}
            & & \textbf{PSNR (dB)} & \textbf{SSIM} & \textbf{PSNR (dB)} & \textbf{SSIM} \\
            \midrule
         Learning-free&   Zero-filled              & 31.00$\pm$3.33       & 0.7848$\pm$0.0616          & 30.86$\pm$1.73       & 0.8065$\pm$0.0376          \\ 
          &  $\ell_1$-Wavelet \cite{cs_mri}        & 25.67$\pm$3.91     & 0.5667$\pm$0.2001      & 28.68$\pm$1.31             & 0.6783$\pm$0.0684                    \\ \midrule
       Diffusion &     CSGM \cite{jalal2021robust}               & 26.52$\pm$3.21         & 0.6789$\pm$0.1220         & -        & -                       \\
          &  ScoreMRI \cite{score_sde}       & 25.72$\pm$1.80                    & 0.5789$\pm$0.0910       & -     & -                       \\
         &   PnP-DM \cite{wu2024principled}    & 26.52$\pm$3.14                  & 0.6383$\pm$0.1320      & -      & -                       \\ \midrule
          End-to-end  &            U-Net \cite{ronneberger2015u}          & 37.07$\pm$2.47       & 0.8803$\pm$0.0504          & 37.27$\pm$1.76       & 0.9211$\pm$0.0272          \\
                 &      {E2E-VN \cite{e2evarnet}}          & 38.33$\pm$3.06       & 0.9048$\pm$0.0732         & 38.06$\pm$2.70       & 0.9620$\pm$0.0107          \\ 
          &  \textbf{Ours}            & {\bf 39.14$\pm$2.93} & {\bf 0.9219$\pm$0.0724} & {\bf 38.82$\pm$2.77} & {\bf 0.9621$\pm$0.0086} \\
            \bottomrule
        \end{tabular}
    }
    \caption{{\bf MRI reconstruction performance on 4$\times$ equispaced} undersampling. NO outperforms existing methods (classical, diffusion, and end-to-end). NO also shows consistent performance across $\K$ space undersampling patterns (Section \ref{sec:krecon}). Zero-filled refers to reconstructing the image from zero-filled $\K$ space.}
        \label{tab:singleres}
\end{table}
\begin{figure*}[t]
\vspace{-1em}
    \centering
    \includegraphics[width=\linewidth]{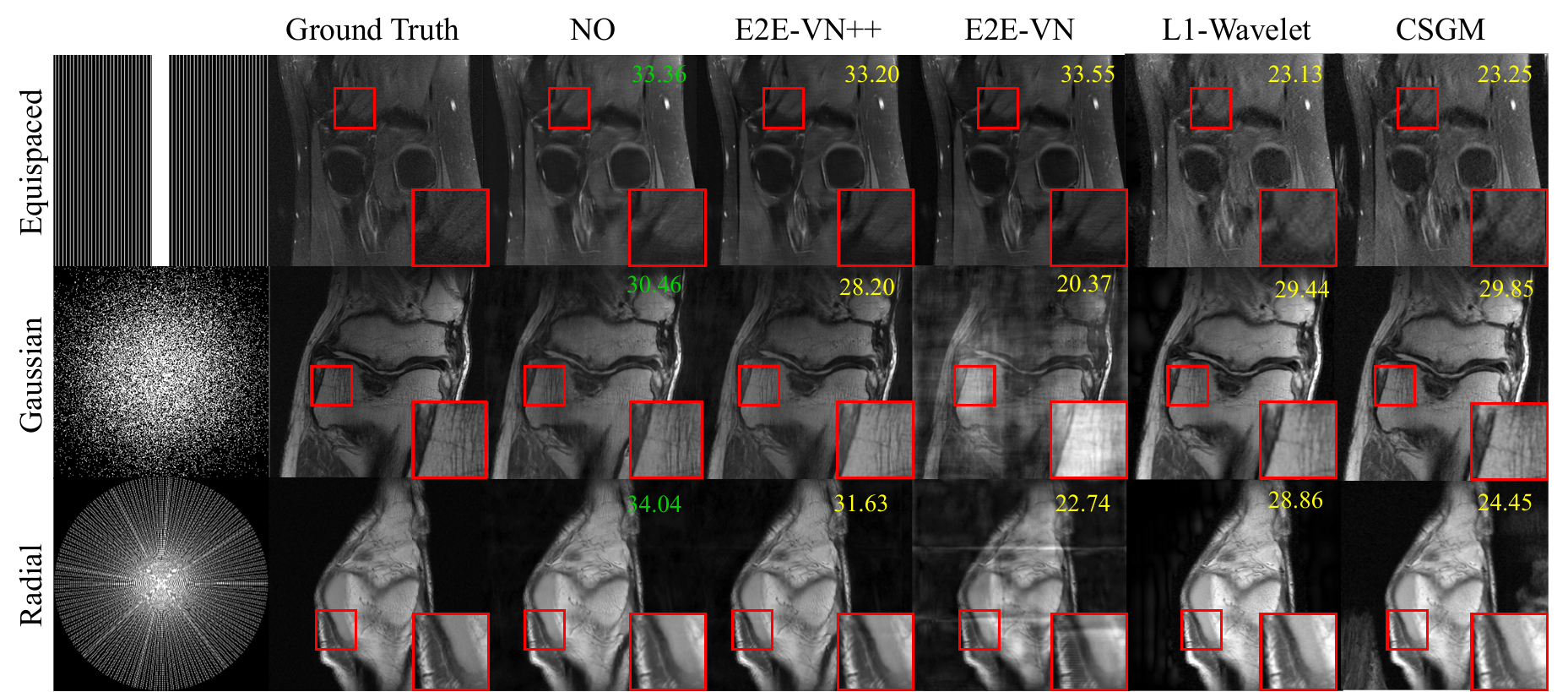}
    \caption{{\bf MRI reconstructions with different undersampling patterns of various methods}: NO (ours), E2E-VN++, E2E-VN \cite{e2evarnet}, L1-Wavelet (learning-free compressed sensing) \cite{cs_mri}, and CSGM (diffusion) \cite{jalal2021robust}.  NO reconstructs high-fidelity images across various downsampling patterns. Zoom-in view in the lower right of each image.  
    \textbf{Row 1:} 4$\times$ Equispaced undersampling. \textbf{Row 2:} 4$\times$ Gaussian 2d undersampling. \textbf{Row 3:} 4$\times$ Radial 2d undersampling.}
    \label{fig:viz_multipatt}
\end{figure*}

\begin{table*}[h]
\centering
\resizebox{\linewidth}{!}{
    \begin{tabular}{clccccccccc}
        \toprule
        & & \multicolumn{3}{c}{\textbf{PSNR (dB) $\uparrow$}} & \multicolumn{3}{c}{\textbf{SSIM $\uparrow$}} & \multicolumn{3}{c}{\textbf{NMSE $\downarrow$}} \\
        \cmidrule(lr){3-5} \cmidrule(lr){6-8} \cmidrule(lr){9-11}
        & \textbf{Pattern} & \textbf{NO (ours)} & \textbf{E2E-VN++} & \textbf{E2E-VN} \cite{e2evarnet} & \textbf{NO (ours)} & \textbf{E2E-VN++} & \textbf{E2E-VN} \cite{e2evarnet} & \textbf{NO (ours)} & \textbf{E2E-VN++} & \textbf{E2E-VN} \cite{e2evarnet} \\
        \midrule
       {\bf In-} & \textbf{Equispaced} & $37.40 \pm 2.61$ & $37.50 \pm 2.79$ & $\mathbf{38.35 \pm 3.05}$ & $0.899 \pm 0.072$ & $0.900 \pm 0.072$ & $\mathbf{0.905 \pm 0.073}$ & $0.007 \pm 0.006$ & $0.007 \pm 0.006$ & $\mathbf{0.006 \pm 0.006}$ \\
       {\bf domain} & \textbf{Random} & $36.66 \pm 2.48$ & $36.79 \pm 2.65$ & $\mathbf{37.34 \pm 2.75}$ & $0.891 \pm 0.070$ & $0.892 \pm 0.072$ & $\mathbf{0.897 \pm 0.071}$ & $0.008 \pm 0.006$ & $0.008 \pm 0.007$ & $\mathbf{0.007 \pm 0.005}$ \\
        & \textbf{Magic} & $\mathbf{38.46 \pm 2.99}$ & $38.34 \pm 3.06$ & $\mathbf{38.94 \pm 3.55}$ & $0.914 \pm 0.070$ & $0.914 \pm 0.069$ & $\mathbf{0.917 \pm 0.071}$ & $\mathbf{0.006 \pm 0.006}$ & $0.007 \pm 0.006$ & $\mathbf{0.006 \pm 0.006}$ \\
        \midrule
        \multirow{3}{*}{\textbf{OOD}} & \textbf{Radial} & $\mathbf{36.23 \pm 2.21}$ & $35.50 \pm 2.24$ & $27.02 \pm 3.92$ & $\mathbf{0.900 \pm 0.071}$ & $0.892 \pm 0.070$ & $0.764 \pm 0.070$ & $\mathbf{0.009 \pm 0.006}$ & $0.011 \pm 0.009$ & $0.069 \pm 0.030$ \\
        & \textbf{Poisson} & $\mathbf{33.42 \pm 2.34}$ & $33.01 \pm 2.67$ & $23.61 \pm 3.85$ & $\mathbf{0.878 \pm 0.062}$ & $0.873 \pm 0.060$ & $0.687 \pm 0.083$ & $\mathbf{0.016 \pm 0.008}$ & $0.017 \pm 0.008$ & $0.152 \pm 0.068$ \\
        & \textbf{Gaussian} & $\mathbf{31.25 \pm 2.70}$ & $30.65 \pm 2.55$ & $23.14 \pm 3.95$ & $\mathbf{0.863 \pm 0.058}$ & $0.851 \pm 0.059$ & $0.673 \pm 0.088$ & $\mathbf{0.024 \pm 0.005}$ & $0.028 \pm 0.007$ & $0.170 \pm 0.073$ \\
        \bottomrule
    \end{tabular}
}
\caption{{\bf MRI reconstruction performance across different undersampling patterns}. Across multiple patterns, NO maintains reconstruction performance, while baselines do not perform well on out-of-domain (OOD) undersampling patterns (Poisson, radial, Gaussian).
Metrics are calculated for the fastMRI knee dataset with a fixed $4\times$ acceleration rate. We observe that the E2E-VN overfits to rectilinear patterns, and drops off heavily when evaluated on the irregular patterns (Poisson, radial, Gaussian).}
\label{tab:k_pattern}
\end{table*}

\begin{figure*}[t]
\vspace{-1em}
    \centering
    \includegraphics[width=\textwidth]{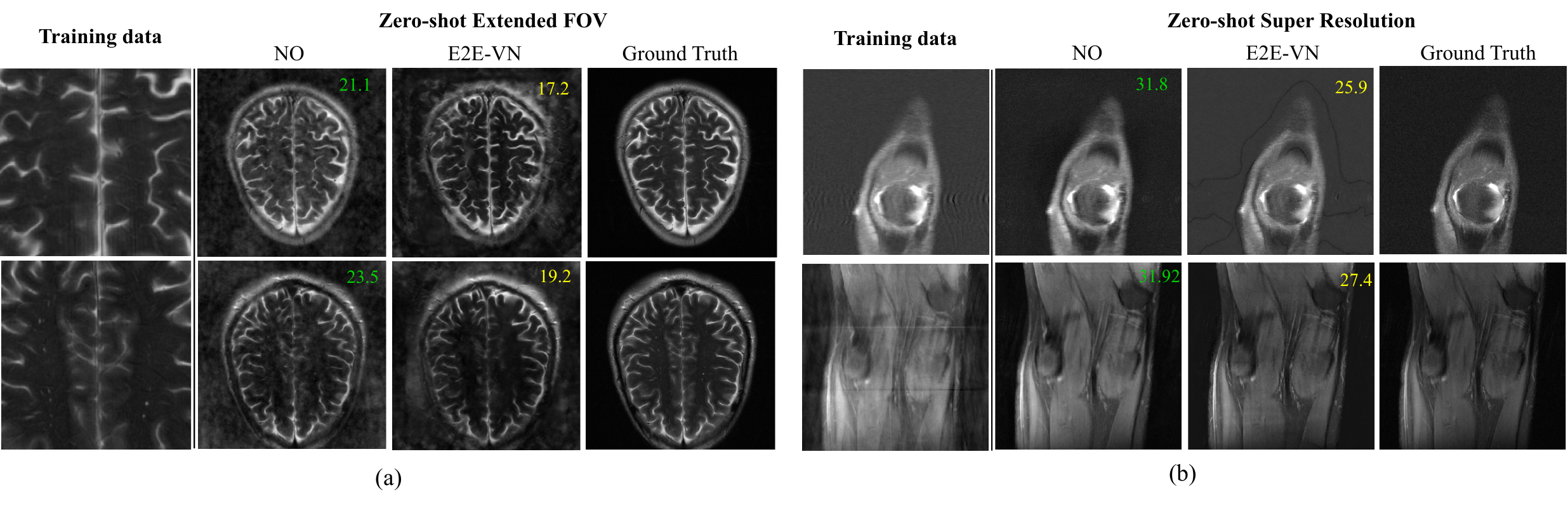}
    \caption{{\bf Zero-shot super-resolution results}  in both extended FOV (\kno) and high-resolution image space (\ino).  \textbf{(a) Zero-shot extended FOV reconstructions}: Our NO model shows fewer artifacts and higher PSNR in the reconstructed brain slices compared to the CNN-based E2E-VN \cite{e2evarnet} on $4\times$ Gaussian, despite neither model seeing data outside the initial $160 \times 160$ FOV during training. \textbf{(b) Zero-shot super-resolution reconstructions in image space} on $2\times$ radial: with input resolution increased to $640 \times 640$ through bilinear interpolation, our NO model preserves reconstruction quality, while E2E-VN \cite{e2evarnet} produces visible artifacts.}%
    \label{fig:superres}
\end{figure*}

\noindent{\bf Neural Operator Model}. NO follows Fig.~\ref{fig:framework}. The \kno ($\K$ space neural operator) and \ino (image space neural operator) are implemented as UDNOs with 2 input and output channels. This is because complex numbers, commonly used in MRI data, are represented using two channels: one for the real part and one for the imaginary part. We provide UDNO details, DISCO kernel basis configurations and training hyper-parameters in Section~\ref{sec:secb} of the Supplementary.

\noindent {\bf Baseline: Compressed Sensing}. We compare with a learning-free compressed sensing method with wavelet $\ell_1$ regularization for a classical comparison \cite{cs_mri}.

\noindent {\bf Baselines: Unrolled Networks}. We compare with the E2E-VN (End-to-End VarNet) \cite{e2evarnet}, which shares a similar network structure with our approach, but uses the standard CNNs with resolution-dependent convolutions. 
Since E2E-VN \cite{e2evarnet} is only trained on specific resolution, we also consider E2E-VN++, where we train \cite{e2evarnet} with multiple-patterns that match our NO's training data for fair comparisons. Number of cascades $t$ is set to 12 following   \cite{e2evarnet}.

\noindent {\bf Baselines: Diffusion}. Diffusion models have shown strong performance on inverse problems such as MRI reconstruction. We compare our approach to three prominent diffusion-based methods that leverage these capabilities: Score-based diffusion models for accelerated MRI (ScoreMRI) \cite{score_sde}, Compressive Sensing using Generative Models (CSGM) \cite{jalal2021robust}, and Plug-and-Play Diffusion Models (PnP-DM) \cite{wu2024principled}. We replicate the experimental settings described in their respective papers. While they report results on MVUE targets, we evaluate metrics on RSS targets at inference for a fair comparison with our methods.

\noindent{\bf Hardware and Training.} While models can be trained on a single RTX 4090 GPU, we accelerate the training of our model and baselines with a  batch size of 16 across 4 A100 (40G) GPUs. We follow baseline settings for comparison.

\noindent {\bf Evaluation Protocols}. We evaluate image reconstruction performance using normalized mean square error (NMSE), peak signal-to-noise ratio (PSNR), and structural similarity index measure (SSIM) which are standard for the fastMRI dataset and MRI \cite{zbontar2018fastMRI}.

\subsection{Reconstruction with Different \\k Space Undersampling Patterns}
We train our NO model, E2E-VN and E2E-VN++ on 4$\times$ \textit{equispaced} samples for 50 epochs. The performance on the single 4$\times$ equispace undersampling pattern in Table \ref{tab:singleres}. We further fine-tune NO and E2E-VN++ for an additional 20 epochs on a small dataset (3,474 samples) of equispaced, random, magic, Gaussian, radial, and Poisson samples.

\noindent {\bf fastMRI Knee}. %
We also provide detailed metric results in Table \ref{tab:k_pattern}, with a line plot in Fig. \ref{fig:lineplot}a, where our NO achieves consistent performance across different patterns. Across all patterns, we achieve an average improvement of 4.17~dB PSNR and 8.4\% SSIM over the E2E-VN. On rectilinear patterns (equispaced, magic, random), our performance remains comparable to E2E-VN++ (0.3 dB PSNR gain). Across the irregular patterns (radial, Gaussian, Poisson), we achieve a 0.6~dB PSNR improvement over the improved baseline (E2E-VN++).%

\noindent {\bf fastMRI Brain}. On irregular patterns, we achieve an average improvement of 4.7~dB PSNR and 10\% SSIM over the E2E-VN. On rectilinear patterns (equispaced, magic, random), our performance remains comparable to the E2E-VN. Detailed numbers are reported in Table \ref{tab:b_pattern} of Supplementary. 

\noindent{\bf Visualization}. We observe visual improvements in reconstruction integrity (see Fig.~\ref{fig:viz_multipatt}). Our model is robust to inference across multiple patterns. We highlight important local regions where our NO is better.

The setting here where multiple patterns are trained together is a common clinical setting where the undersampling patterns are known. We also consider the setting where undersampling patterns are unknown. Zero-shot evaluations of the equispaced-trained (4$\times$) model across different patterns show that our NO shows 1.8~dB PSNR gain over E2E-VN.

\subsection{Reconstruction with Different \\k Space Undersampling Rates}%
\label{sec:krecon}
We train our NO model, E2E-VN and E2E-VN++ on 4$\times$ \textit{equispaced} samples for 50 epochs. We further fine-tune NO and E2E-VN++ for an additional 20 epochs on a small dataset (3,474 samples) of 4$\times$, 6$\times$, 8$\times$, and 16$\times$ equispaced samples.

For {\bf fastMRI Knee}, we report the multi-rate performance in Fig. \ref{fig:lineplot}b and Table \ref{tab:k_rate} of the Supplementary. For {\bf fastMRI Brain}, we report the multi-rate performance in Table \ref{tab:b_rate} of the Supplementary.
Our neural operator model consistently outperforms the E2E-VN \cite{e2evarnet}, achieving 3.2~dB higher PSNR and 5.8\% higher SSIM on fastMRI knee and 2.0~dB higher PSNR and 7.5\% higher SSIM on fastMRI brain. %

\begin{figure}[t]
    \centering
    \includegraphics[width=1\columnwidth]{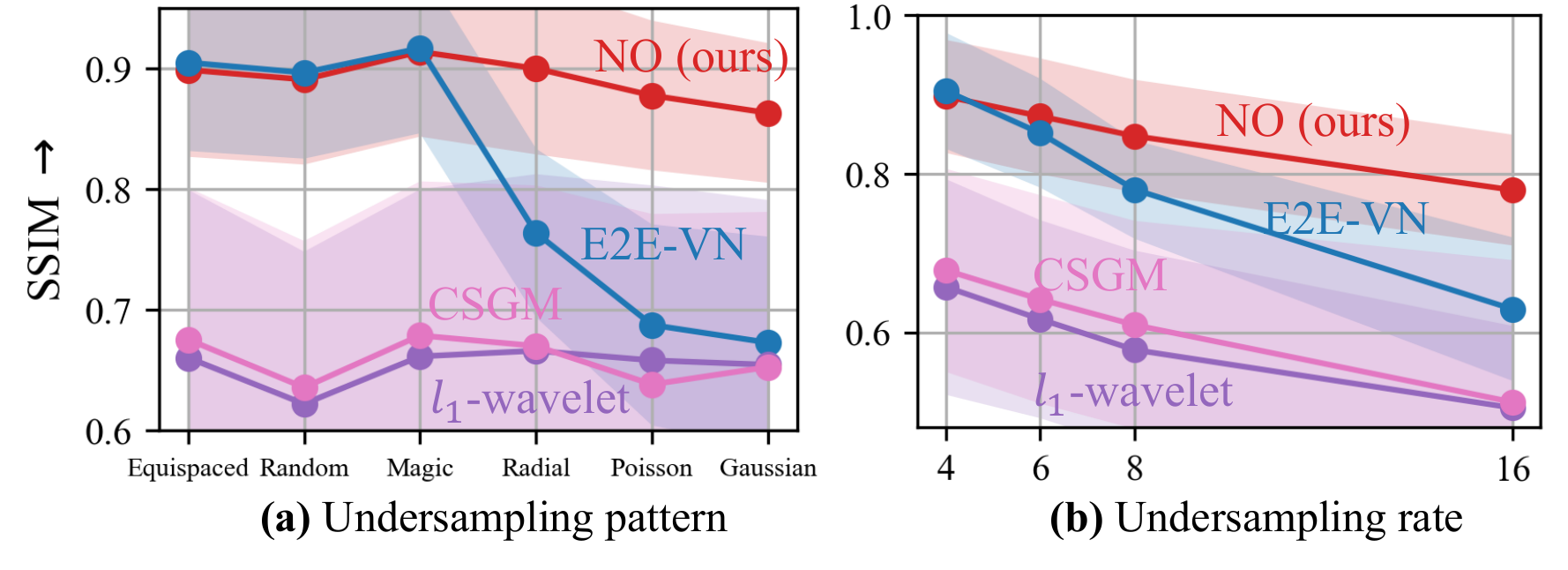}
    \caption{{\bf Performance across different undersampling patterns and rates} of ours and baseline methods: end-to-end \cite{e2evarnet}, diffusion \cite{jalal2021robust} and learning-free \cite{cs_mri}.
    Our NO  remains relatively consistent in performance when evaluated at different undersampling patterns and rates. Note that a high undersampling rate makes the task more difficult and thus a worse score is expected.}%
    \label{fig:lineplot}
\end{figure}

    \subsection{Zero-Shot Super-Resolution}
    \label{sec:superres}
    
    We study \ino and \kno zero-shot super-resolution performance and compare them with E2E-VN \cite{e2evarnet}.
    
    \noindent{\bf Higher MRI Resolution} with \ino super-resolution. We train our NO model and the E2E-VN models on $320\times320$ knee samples. We then keep the \kno unchanged and use bilinear interpolation to increase the input to \ino to $640\times640$. We directly evaluate models without fine-tuning against fully sampled $640 \times 640$ bilinear interpolated ground truth reconstructions. For \cite{e2evarnet} relying on CNNs, the absolute kernel size stays the same, and the ratio of kernel size over feature map is halved, while the ratio of NO stays the same. 
    Compared to our NO model, the CNN-based E2E-VN \cite{e2evarnet} produces reconstructions with noticeable artifacts and higher PSNR and image reconstruction quality (Fig.~\ref{fig:intro}d and Fig.~\ref{fig:superres}b). %

    \noindent{\bf Larger MRI FOV} with \kno super-resolution.
$\K$ space super-resolution expands the MRI reconstruction field of view (FOV). To validate model performance, we design a proof-of-concept FOV experiment. Our NO model and the E2E-VN \cite{e2evarnet} train on $160 \times 160$ downsampled $\K$ space brain slice samples, where sparse $\K$ space sampling results in a reduced FOV in image space. We then perform zero-shot inference on $320 \times 320$ full-FOV $\K$ space data. Although neither model encounters data outside the $160 \times 160$ FOV during training, our NO model reconstructs features in this extended region with significantly fewer artifacts compared to E2E-VN \cite{e2evarnet} (visualizations in Fig.~\ref{fig:superres}a).

\subsection{Additional Analysis}%
\label{sec:analysis}

\begin{table}[htb]
\centering
\resizebox{\linewidth}{!}{
    \begin{tabular}{llcc}
        \toprule
        {\bf Category} &\textbf{Method} & \textbf{Inference Time (s)} & \textbf{Tuning$^*$ Required} \\
        \midrule
       Learning-free& $\ell_1$-Wavelet \cite{cs_mri} & 5.45 & \cmark  \\ \midrule
      Diffusion&  CSGM  \cite{jalal2021robust} & 93.84 & \cmark\\ 
       & PnP-DM \cite{wu2024principled}& 84.46 & \cmark\\
       & ScoreMRI \cite{score_sde} & 96.16 & \cmark \\ \midrule
      Variational &  E2E-VN \cite{e2evarnet} & 0.104 & \xmark \\ 
        & {NO (ours)} & 0.158 & \xmark \\
        \bottomrule
    \end{tabular}
}
\caption{{\bf Inference and tuning time of methods} tested on NVIDIA A100.   NO  is approximately $600\times$ faster than diffusion, and $35\times$ faster than the classical baseline based on learning-free compressed sensing methods. *Tuning refers to the $\K$ undersampling pattern-specific hyperparameter tuning during inference/after model training. Both the $\ell_1$-Wavelet \cite{cs_mri} ($~\sim0.5$ hrs per pattern) and diffusion methods ($~\sim 6$ hrs per pattern) require pattern-specific tuning, while our NO is trained once for all patterns.} \label{tab:times}
\vspace{-1em}
\end{table}

\noindent{\bf Model Inference and Tuning Time}. In Table~\ref{tab:times}, we compare the model development and inference times of our end-to-end neural operator (NO) with diffusion models. We observe that diffusion models require pattern-specific hyper-parameter tuning and are over 600 times slower in inference. 
MRI-diffusion models \cite{jalal2021robust,wu2024principled,score_sde} are unconditionally trained and undersampling patterns are not available during training. Thus, we empirically tune hyperparameters such as learning rate and guidance scale for each downsampling pattern for approximately 6 hours each time.
 Traditional learning-free methods like $\ell_1$-Wavelet \cite{cs_mri} still require hyperparameter tuning for specific $\K$ undersampling patterns during optimization.
Consequently, end-to-end methods, e.g. NO, are significantly more efficient.

\noindent {\bf Performance Under Same Parameter Size}. We show our NO outperforms baseline unrolled network E2E-VN \cite{e2evarnet} on different patterns and rates with a similar architecture and number of parameters in the Supplementary.

\section{Conclusion}

Our unified model for compressed sensing MRI addresses the need to train multiple models for different measurement undersampling patterns and image resolutions, a common clinical issue. By leveraging discretization-agnostic neural operators, the model captures both local and global features, enabling flexible MRI reconstruction. With extensive experiments on fastMRI knee and brain datasets, our model maintains consistent performance across undersampling patterns and outperforms state-of-the-art methods in accuracy and robustness. It also enhances zero-shot super-resolution and extended FOV (field of view). %
The work has some limitations: {\bf 1)} We only explore one neural operator design, DISCO, and future work could explore other operator learning architectures for MRI. {\bf 2)} We only benchmark the image reconstruction performance without diagnostic accuracy, which is of more clinical relevance.

In short, our approach offers a versatile solution for efficient MRI, with significant utility in clinical settings where flexibility and adaptability to varying undersampling patterns and image resolutions are crucial.

\newpage{}
\section*{Acknowledgment}
This work is supported in part by ONR (MURI grant N000142312654 and N000142012786). J.W. is supported in part by the Pritzker AI+Science initiative and Schmidt Sciences. A.S.J. and A.C. are supported in part by the Undergraduate Research Fellowships (SURF) at Caltech. Z.W. is supported in part by the Amazon AI4Science Fellowship. B.T. is supported in part by the Swartz Foundation Fellowship. M.L.-S. is supported in part by the Mellon Mays Undergraduate Fellowship. A.A. is supported in part by Bren endowed chair and the AI2050 senior fellow program at Schmidt Sciences.
{
    \small
    \bibliographystyle{ieeenat_fullname}
    \bibliography{ref.bib}
}

\clearpage
\newpage
\newpage
\appendix
\clearpage
\maketitlesupplementary

In the supplementary, we first present more details of the proposed U-shaped DISCO Neural Operator (UDNO, in Section \ref{sec:udno_a}), a main building block of \ino and \kno of our framework. We then provide more details of the machine learning framework implementation (Section \ref{sec:secb}) as well as additional numerical results of the multi-pattern and multi-rate undersampling experiments (Section \ref{sec:a_undersampling}).
In Section \ref{sec:ablation} we include additional ablation and analysis, on comparing CNN and NO kernels and their performance under the same parameter size, followed by details about DISCO and the justification of its basis choice in Section \ref{sec:discodetails}.

\section{UDNO Architecture}
\label{sec:udno_a}
\begin{figure*}[ht]
    \centering
    \includegraphics[width=0.85\linewidth]{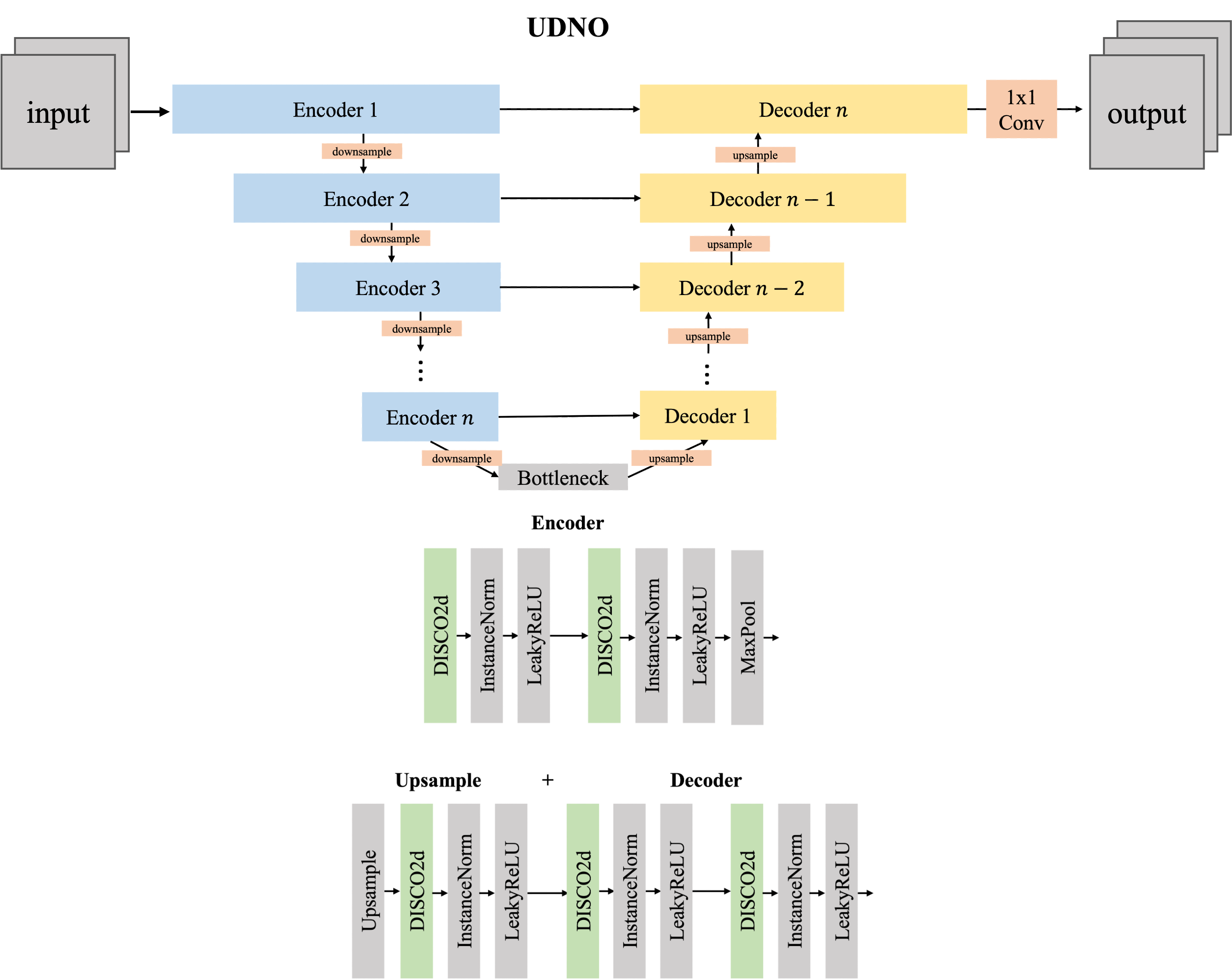} 
    \caption{ {\bf UDNO architecture}. We propose a U-shaped neural operator (UDNO) to capture multi-scale features of the input. The UDNO uses discrete-continuous convolutions (DISCOs)~\citep{ocampo2022} as the local integral operator. The final 1x1 convolution allows the module to flexibly project to the desired number of output channels and is resolution invariant by virtue of being a pointwise operation. The UDNO is an end-to-end \textit{neural operator}.}
    \label{fig:udnoa}
\end{figure*}

The motivation behind the U-shaped architecture is to capture multi-scale features by integrating high-level contextual information with low-level details. Its encoder-decoder structure, enhanced by skip connections, enables precise localization of features even with limited annotated data. In our approach, we extend this idea through UDNO, which is applied to both the physical and frequency domains for MRI reconstruction—unlike methods such as FNO \cite{fnop} used for PDE data that incorporate a frequency cut. This difference arises because PDE data typically comes from smooth functions, where low frequencies are dominant and high frequencies mainly represent noise. In contrast, imaging data benefits from retaining both low-frequency information and high-frequency details (e.g., edges).

We provide additional details of the proposed UDNO (U-Shaped DISCO Neural Operator) architecture. Fig. \ref{fig:udnoa} depicts the overall architecture, which mimics the U-Net \cite{ronneberger2015u}. We use the updated implementation of the U-Net in \cite{e2evarnet}. Our network architecture has two differences. First, all traditional convolutions are replaced with their DISCO counterparts. Second, transpose convolutions are replaced by an interpolation upsampling step, followed by a DISCO2d layer, InstanceNorm layer, and LeakyReLU activation. 
DISCO2d layers function as drop-in replacements for traditional 2d convolution layers. They do not change the spatial dimension of the input. The UDNO is an end-to-end \textit{neural operator}.

As in the traditional U-Net \cite{ronneberger2015u}, each encoder block halves the spatial dimensions and doubles the feature channels. Each decoder step (upsampling + decoder) doubles the spatial dimensions and halves the feature channels. Skip connections are included, as in the original architecture. All components of the UDNO operate in the function space and are not tied to a specific discretization, thus making the model an \textit{end-to-end} neural operator.

\begin{figure}[t!]
    \centering
    \includegraphics[width=1\linewidth]{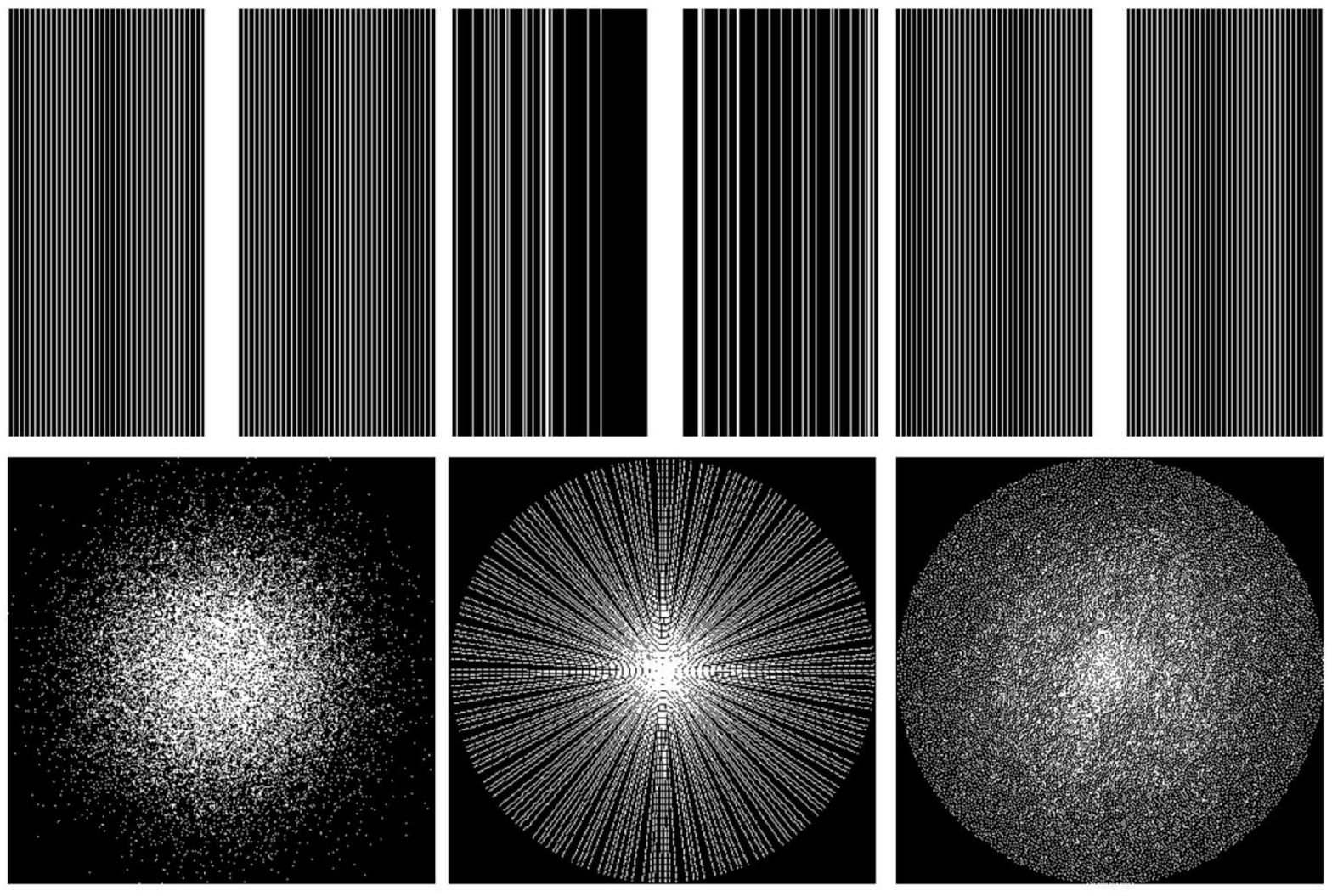}
    \caption{{\bf Undersampling mask patterns}. The visualized patterns are all for the 4$\times$ acceleration rate. \textbf{Top:} Rectilinear patterns: Equispaced, Random, Magic. \textbf{Bottom:} Irregular patterns: Gaussian, Radial, Poisson.}
    \label{fig:mask_fns}
    \vspace{-1em}
\end{figure}

\section{Additional Implementation Details}
\label{sec:secb} 

\subsection{Undersampling Configurations}
We summarize the configurations of different CS-MRI undersampling rates in Table \ref{tab:aliases} and undersampling patterns in Fig. \ref{fig:mask_fns}.

\subsection{Learning Sensitivity Maps for Multi-Coil MRI}

In MRI reconstruction, the sensitivity map $S_i$ for the $i^{\text{th}}$ coil is needed for coil reductions and expansions. Inspired by \cite{e2evarnet}, we use a UDNO with 4 encoder/decoder steps, 8 hidden channels, $0.02$ DISCO radius (assuming the domain is $[-1,1]^2$), and the kernel basis from \cite{local_no} with 1 isotropic basis and 5 anisotropic basis rings, each containing 7 basis functions. We use this UDNO to predict the sensitivity map $S_i$ from the input coil measurement $\K_i$. We then follow \cite{e2evarnet} to combine multiple coils weighted by the corresponding learned sensitivity maps.

\subsection{UDNO and DISCO Implementation Details}
\label{sec:udno_disco_implementation}

Both \kno and \ino use DISCO layers using the \textit{linear-piecewise} kernel basis from \cite{local_no} with 1 isotropic basis and 5 anisotropic basis rings, each containing 7 basis functions. The \kno (measurement space neural operator) is implemented as a UDNO with $2$ input and output channels, $16$ hidden channels, and $4$ depth (encoder/decoder steps). \kno DISCO \ino have a radius cutoff of $0.02$. The \ino (image-space neural operator) is implemented as a UDNO with $2$ input and output channels, $18$ hidden channels, and $4$ encoder/decoder steps. \ino DISCO kernels have a radius cutoff of $0.02$ with the same internal basis shape. We train both our model and the baseline with SSIM loss, and $0.0003$ learning rate.

To compare the choice of basis function (piecewise linear, Zernike, and Morlet), we train our neural operator with a single cascade on a 30\% subset of the fastMRI knee dataset for 15 epochs. We find that empirically, the piecewise linear basis outperforms both the Zernike and Morlet bases by at least $3$ PSNR. All kernels have a similar number of parameters. Results are provided in Table \ref{tab:kernel_basis}.

\begin{table}[t!]
\centering
\resizebox{0.85\linewidth}{!}{
\begin{tabular}{lccc}
\toprule
\textbf{Kernel Basis} & \textbf{PSNR $\uparrow$} & \textbf{SSIM $\uparrow$} & \textbf{NMSE $\downarrow$} \\
\midrule
\textbf{Piecewise Linear} & $\mathbf{36.125}$ & $\mathbf{0.884}$ & $\mathbf{0.009}$\\
\textbf{Zernike} & $32.983$ & $0.838$ & $0.017$\\
\textbf{Morlet} & $32.755$ & $0.835$ & $0.018$ \\
\bottomrule
\end{tabular}
}
\caption{\textbf{Kernel basis experiment results.} We train our neural operator model with the piecewise-linear, Zernike, and Morlet bases, comparing empirical reconstruction results. The Piecewise Linear basis outperforms both the Zernike and Morlet by at least 3 PSNR.}
\label{tab:kernel_basis}
\end{table}

\begin{table}[t!]
  \centering
 \resizebox{0.85\linewidth}{!}{\begin{tabular}{lcc}
    \toprule
    \textbf{Alias} & \textbf{Acceleration rate} & \textbf{Center fraction rate} \\
    \midrule
    16$\times$ & 16 & 0.02 \\
    8$\times$ & 8 & 0.04 \\
    6$\times$ & 6 & 0.06 \\
    4$\times$ & 4 & 0.08 \\
    \bottomrule
  \end{tabular}
  }
    \caption{
  $\K$ {\bf space undersampling configurations} (acceleration and center fraction parameters) used for MRI experiments. We follow the \cite{e2evarnet} and \cite{e2evarnet}}
    \label{tab:aliases}
\end{table}

\begin{table}[t]
\centering
\resizebox{\linewidth}{!}{
\begin{tabular}{lcccc}
\toprule
& \multicolumn{2}{c}{\textbf{NO (ours)}} & \multicolumn{2}{c}{\textbf{E2E-VN~\cite{e2evarnet}}} \\
\cmidrule(lr){2-3} \cmidrule(lr){4-5}
\textbf{Rate} & \textbf{PSNR $\uparrow$} & \textbf{SSIM $\uparrow$} & \textbf{PSNR $\uparrow$} & \textbf{SSIM $\uparrow$} \\
\midrule
\textbf{4$\times$} & $37.215 \pm 2.466$ & $0.897 \pm 0.071$ &  $\mathbf{38.329 \pm 3.062}$ & $\mathbf{0.905 \pm 0.073}$ \\
\textbf{6$\times$} & $\mathbf{35.452 \pm 2.150}$ & $\mathbf{0.872 \pm 0.073}$ & $32.770 \pm 2.064$ & $0.851 \pm 0.069$ \\
\textbf{8$\times$} & $\mathbf{33.598 \pm 1.892}$ & $\mathbf{0.848 \pm 0.071}$ & $28.346 \pm 2.407$ & $0.780 \pm 0.062$ \\
\textbf{16$\times$} & $\mathbf{29.241 \pm 2.402}$ & $\mathbf{0.779 \pm 0.070}$ & $23.181 \pm 3.558$ & $0.629 \pm 0.090$ \\
\bottomrule
\end{tabular}
}
\caption{{\bf fastMRI Knee performance across different undersampling rates.} %
We compare our NO model's knee reconstruction performance to the E2E-VN \cite{e2evarnet}, assessing for robustness against different undersampling rates. Both models are trained on equispaced 4$\times$ knee samples, and evaluated across 4$\times$, 6$\times$, 8$\times$, and 16$\times$ equispaced validation samples. Notice that over the irregular patterns, our model shows an increase of 3.22 dB PSNR and 5.8\% SSIM.}
\label{tab:k_rate}
\end{table}

\begin{table}[t]
\centering
\resizebox{\linewidth}{!}{
\begin{tabular}{lcccc}
\toprule
& \multicolumn{2}{c}{\textbf{NO (ours)}} & \multicolumn{2}{c}{\textbf{E2E-VN~\cite{e2evarnet}}} \\
\cmidrule(lr){2-3} \cmidrule(lr){4-5}
\textbf{Pattern} & \textbf{PSNR $\uparrow$} & \textbf{SSIM $\uparrow$} & \textbf{PSNR $\uparrow$} & \textbf{SSIM $\uparrow$} \\
\midrule
\textbf{Equispaced} & $37.106 \pm 1.646$ & $0.952 \pm 0.010$ & $38.063 \pm 2.701$ & $0.962 \pm 0.011$ \\
\textbf{Random} & $36.051 \pm 1.665$ & $0.945 \pm 0.015$ & $37.025 \pm 2.187$ & $0.957 \pm 0.010$ \\
\textbf{Magic} & $38.270 \pm 1.985$ & $0.960 \pm 0.011$ & $38.463 \pm 2.967$ & $0.965 \pm 0.011$ \\
\textbf{Radial} & $36.498 \pm 1.792$ & $0.948 \pm 0.015$ & $25.225 \pm 2.126$ & $0.722 \pm 0.063$ \\
\textbf{Poisson} & $33.936 \pm 2.047$ & $0.924 \pm 0.020$ & $22.117 \pm 1.487$ & $0.670 \pm 0.046$ \\
\textbf{Gaussian} & $32.725 \pm 2.004$ & $0.910 \pm 0.018$ & $25.283 \pm 3.336$ & $0.730 \pm 0.098$ \\
\bottomrule
\end{tabular}
}
\caption{{\bf fastMRI Brain performance across different undersampling patterns.} We compare our NO model's brain reconstruction performance to the E2E-VN \cite{e2evarnet}, assessing for robustness against different undersampling patterns. Both models are trained on equispaced 4$\times$ brain samples, and evaluated across multiple patterns. Notice that over the irregular patterns, our model shows a significant 10 dB PSNR and 22\% SSIM improvement on average. Our NO model is robust to different patterns, while the E2E-VN overfits to the rectilinear patterns (equispaced, random, magic).}
\label{tab:b_pattern}
\end{table}

\begin{table}[t]
\centering
\resizebox{\linewidth}{!}{
\begin{tabular}{lcccc}
\toprule
& \multicolumn{2}{c}{\textbf{NO (ours)}} & \multicolumn{2}{c}{\textbf{E2E-VN~\cite{e2evarnet}}} \\
\cmidrule(lr){2-3} \cmidrule(lr){4-5}
\textbf{Rate} & \textbf{PSNR $\uparrow$} & \textbf{SSIM $\uparrow$} & \textbf{PSNR $\uparrow$} & \textbf{SSIM $\uparrow$} \\
\midrule
\textbf{4$\times$} & $36.851 \pm 2.334$ & $0.952 \pm 0.027$ & $38.294 \pm 3.030$ & $0.959 \pm 0.027$ \\
\textbf{6$\times$} & $34.575 \pm 2.404$ & $0.934 \pm 0.030$ & $33.418 \pm 2.675$ & $0.925 \pm 0.030$ \\
\textbf{8$\times$} & $32.246 \pm 2.306$ & $0.912 \pm 0.031$ & $28.827 \pm 3.197$ & $0.856 \pm 0.047$ \\
\textbf{16$\times$} & $27.561 \pm 2.620$ & $0.836 \pm 0.044$ & $22.694 \pm 3.097$ & $0.594 \pm 0.104$ \\
\bottomrule
\end{tabular}
}
\caption{{\bf fastMRI Brain performance across different undersampling rates.} Comparisons of the reconstruction quality of our NO model with the E2E-VN \cite{e2evarnet} across various undersampling rates demonstrate that our model maintains robustness at higher undersampling rates and the E2E-VN shows significant degradation in both metrics, particularly at extreme undersampling (e.g., 16$\times$).}
\label{tab:b_rate}
\end{table}

\subsection{Baseline Hyperparameter Search Details}
For the diffusion baseline CSGM, we tuned \texttt{step\_lr} and \texttt{mse} parameters in their \href{https://github.com/utcsilab/csgm-mri-langevin/blob/main/main.py}{official github repo}) 
using Bayesian optimization. The search algorithm was run on 6 representative images outside of the test set for around 50 iterations with the search space defined in Table \ref{tab:search_space}. 
 For  E2E-VN baselines, we tune the number of layers in each cascade, learning rate  and schedule.
\begin{table}[h]
\centering
\small
\resizebox{\columnwidth}{!}{ %
    \begin{tabular}{l|ccc} \toprule
        \textbf{Parameter} & \textbf{Lower bound} & \textbf{Upper bound} & \textbf{Distribution} \\
        \midrule
        \texttt{step\_lr} &$10^{-5}$ & $10^{-4}$ & Log-uniform \\
        \texttt{mse} & $0$ & $5$ & Uniform \\
        \bottomrule
    \end{tabular}
}
\caption{Hyperparameter search space for the diffusion baseline.}
\label{tab:search_space}
\end{table}

\section{Additional Results Across Undersampling Patterns and Rates}
\label{sec:a_undersampling}
We summarize the numerical results of the performance of the proposed neural operator (NO) and the End-to-End VarNet baseline \cite{e2evarnet} across different undersampling patterns and rates on the fastMRI \cite{zbontar2018fastMRI} knee and brain dataset.

\paragraph{fastMRI Knee.} Results for multiple patterns are in Table \ref{tab:k_pattern} of the paper and those for multiple rates are in Table \ref{tab:k_rate}.

\paragraph{fastMRI Brain.} Results for multiple patterns are in Table \ref{tab:b_pattern} and those for multiple rates are in Table \ref{tab:b_rate}.

\begin{figure}[t]
    \centering
    \includegraphics[width=1\columnwidth]{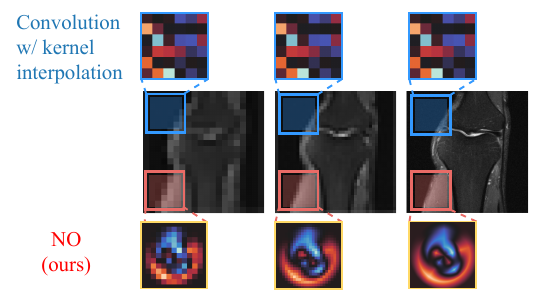}
    \caption{{\bf Ablation study: consistent kernel size to image size ratio for both CNNs and NOs}. As illustrated in Fig.~\ref{fig:intro}b, CNNs have inconsistent relative kernel size when image resolution changes. In this ablation study, we manually resize the CNN kernel with bilinear interpolation to make its relative kernel size consistent for different resolutions and compare the performance with the NO.
    }%
    \label{fig:k_int}
\end{figure}

\section{Additional Ablation Studies and Analysis}
\label{sec:ablation}
\noindent {\bf Rescaling CNN Kernel Size for Consistent Ratio}. %
 As illustrated in Fig. \ref{fig:intro}b, CNNs have inconsistent kernel size to image size ratio when image resolution changes. 
 We want to compare  NO kernels, parameterized in the function space, with CNN kernels by eliminating the factor of kernel size ratio with CNN kernel interpolation. 
 In this ablation study, we manually resize the CNN kernel in \cite{e2evarnet} with bilinear interpolation to make its relative kernel size consistent for different resolutions and call it E2E-VN-INTERP. We compare its performance with the NO.
Specifically, in a super-resolution experiment as follows, we train both our NO and the E2E-VN-INTERP  on $320\times320$ equispaced 4$\times$ knee samples, with a similar setting as in Section \ref{sec:superres} (\ino MRI higher-resolution experiment). Then, we perform zero-shot inference on higher resolution $640\times640$ samples in image space. 

\begin{figure}
    \centering
    \includegraphics[width=\columnwidth]{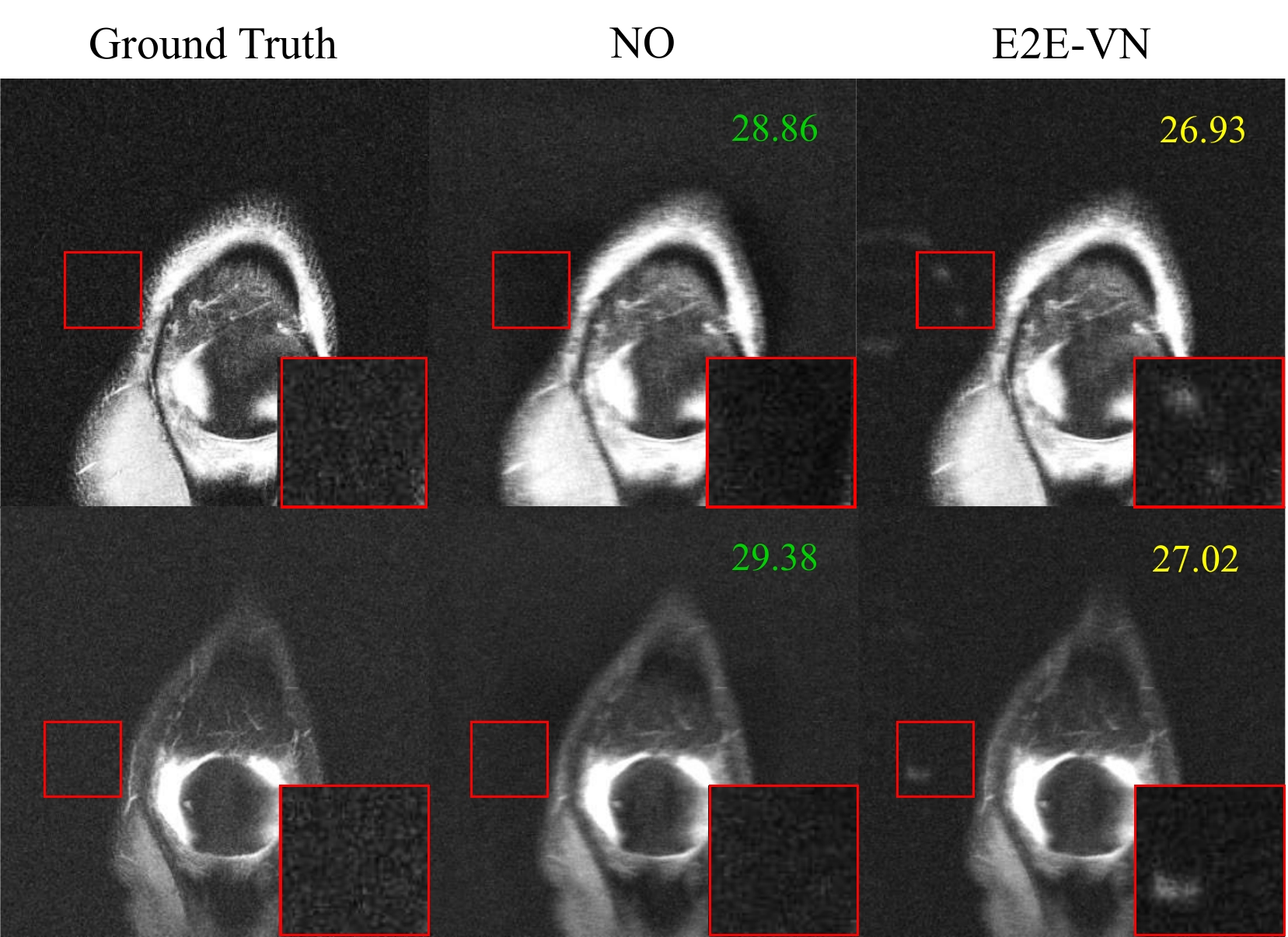}
    \caption{\textbf{Zero-shot inference on higher-resolution samples (NO vs. E2E-VN with interpolated kernels).} While both models are able to recover overall structure, notably, the E2E-VN suffers from hallucinations and noise artifacts in the area surrounding the subject's knee.}
    \label{fig:kinterp}
\end{figure}

Our NO model leverages DISCO convolutions, which enable zero-shot inference on arbitrary resolutions, making them inherently resolution-agnostic (Fig.~\ref{fig:intro}a). In contrast, traditional CNN kernels are designed for fixed resolutions. For instance, the original $3\times3$ kernels of the E2E-VN model, backed up by CNNs, cannot directly scale to the larger $640\times640$ inference resolution. One approach to address this is by resizing the learned kernels to $6\times6$ with bilinear interpolation while preserving their norms, as we follow \cite{local_no} and use quadrature weights to perform the integration. We adopt this method, comparing our NO model with the kernel-scaled E2E-VN-INTERP model. Side-by-side visualization results are presented in Fig.~\ref{fig:kinterp}, where we observe a slightly worse reconstruction performance in the background region of E2E-VN-INTERP compared to VN. Also, E2E-VN-INTERP outperforms the E2E-VN with inconsistent kernel size, validating the need to keep a consistent relative kernel size.  

\begin{figure}[]
    \includegraphics[width=0.9\columnwidth]{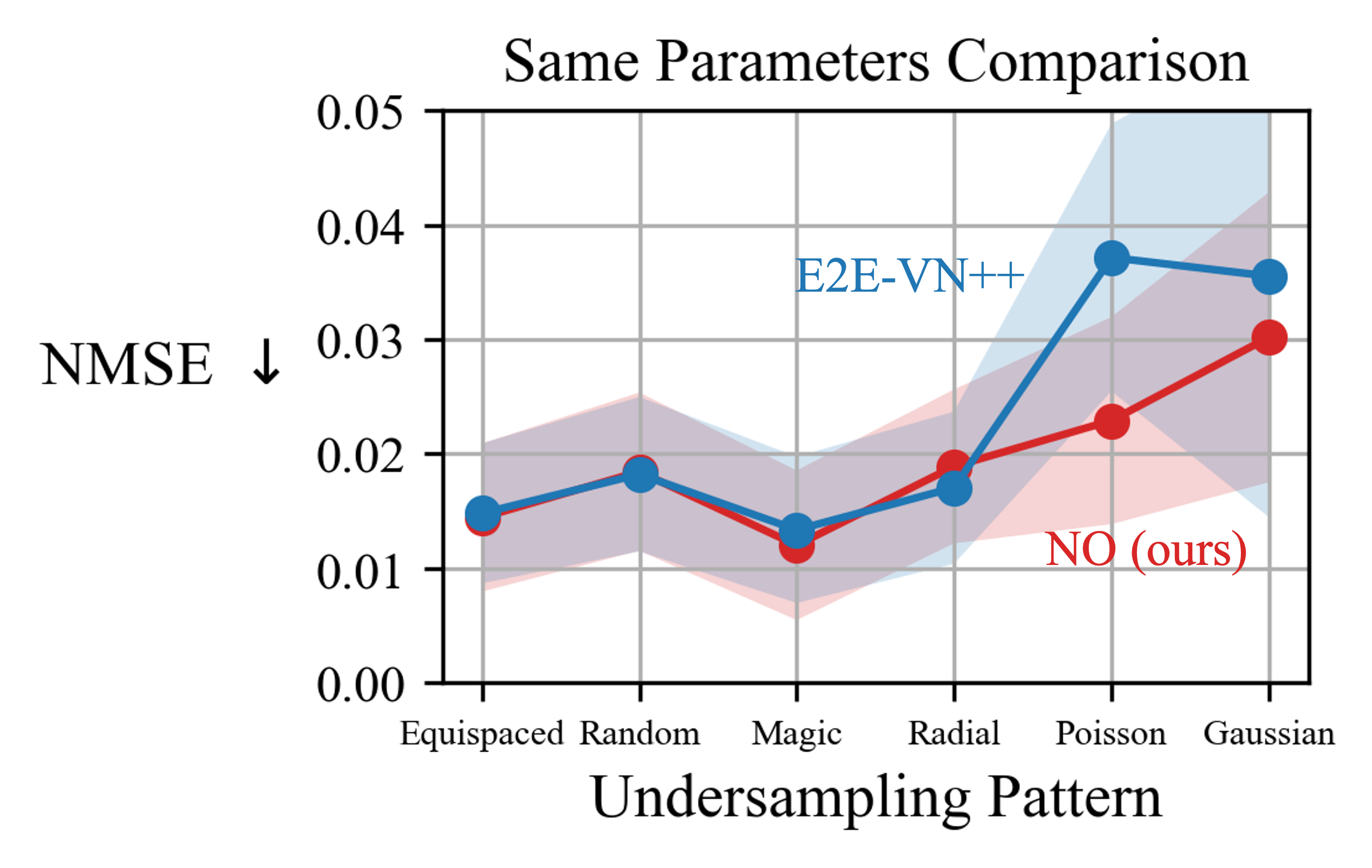}
    \caption{\textbf{Comparison between same parameter (21.7M) NO and E2E-VN++ with NMSE ($\downarrow$)}. While performance is similar on rectilinear patterns, on irregular patterns our NO model achieves lower NMSE than the E2E-VN of same size. On the Poisson undersampling pattern, we achieve 45\% lowering NMSE. On the Gaussian undersampling pattern, we achieve 15\% lower NMSE. We also notice that our NO model exhibits lower variance in its prediction performance.}
    \label{fig:sameparam}
\end{figure}

\noindent {\bf Performance Under Same Parameter Size.} Additionally, we conduct an experiment comparing the NO and E2E-VN models, ensuring both have an identical number of trainable parameters (21.7M). Both models are pretrained on 4$\times$ equispaced fastMRI brain samples for 10 epochs. Then, both are trained for an additional epoch, in which they see samples from all patterns together. We plot cross-pattern performance in Fig. \ref{fig:sameparam}.

\vfill 

\begin{figure}[h!]
    \centering
    \includegraphics[width=0.9\linewidth]{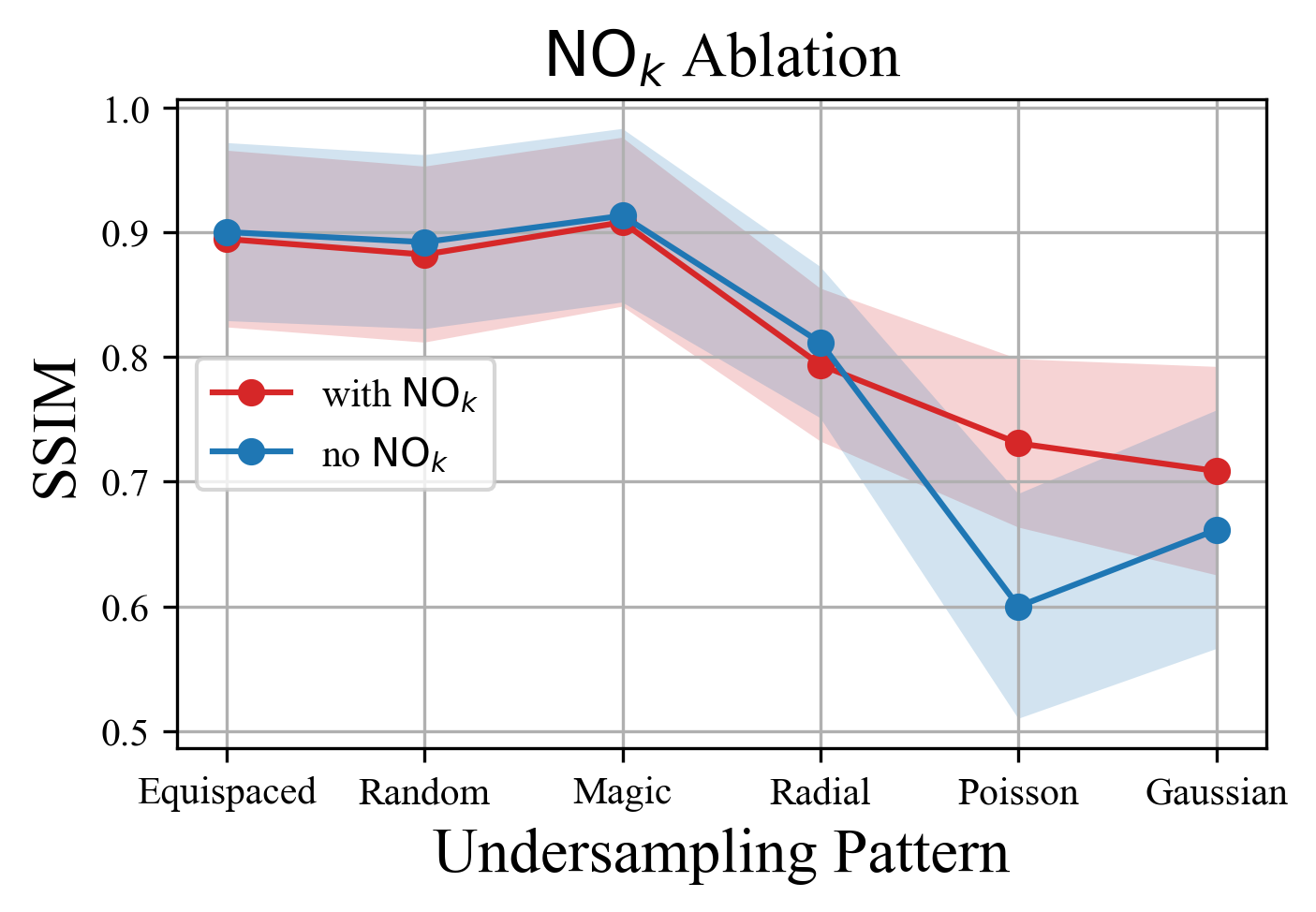}
    \label{fig:enter-label}
    \caption{The ablation study of \kno.}
        \label{fig:knoa}
\end{figure}

\noindent {\bf Functions of \kno.} We perform an ablation study of our \kno module, training both models on a small subset of the full 4$\times$ equispaced training set and plot zero-shot SSIM scores across all patterns. The \kno increases zero-shot SSIM by 5.3\% across irregular patterns (Fig. \ref{fig:knoa}).

\noindent {\bf FLOPs of models.} We measure the number of forward passes and GFLOPs required in a single inference in Table \ref{tab:flops}. Notice that diffusion requires multiple forward passes for a single inference, which is why the computational cost is several order of magnitudes greater.

\begin{table}[h]
\centering
\small
\resizebox{\columnwidth}{!}{ %
    \begin{tabular}{l|cccc} \toprule
        \textbf{Metric} & \textbf{NO (ours)} & \textbf{CSGM} & \textbf{ScoreMRI} & \textbf{PnP-DM} \\
        \midrule
        \# forward passes & 1 & 3465 & 4000 & 2651 \\
        GFLOPs & 171 & 823K & 950K & 630K \\
        \bottomrule
    \end{tabular}
}
\caption{Comparison of GFLOPs and forward passes required per method.}
\label{tab:flops}
\end{table}

\section{DISCO: Discrete-Continuous Convolutions}
\label{sec:discodetails}
\subsection{Definition}

Discrete-continuous (DISCO) convolutions \cite{ocampo2022} generalize the standard (continuous) convolution to Lie groups and quotient spaces. The approach is inspired by conventional convolutional layers, which efficiently implement local operations in neural networks but—upon grid refinement—converge to pointwise linear operators.

\begin{definition}[Group Convolution]
Let $\kappa,v: G \to \mathbb{R}$ be functions on a group $G$. Their convolution is defined as
\begin{equation}\label{eq:group_conv}
(\kappa \star v)(g) = \int_G \kappa(g^{-1}x)\, v(x)\, \mathrm{d}\mu(x),
\end{equation}
with $g,x\in G$ and $\mathrm{d}\mu(x)$ the invariant Haar measure.
\end{definition}

\begin{definition}[DISCO Convolutions]
Given a quadrature rule with points $x_j\in G$ and weights $q_j$, the convolution \eqref{eq:group_conv} is approximated by
\begin{align}
(\kappa \star v)(g) \approx \sum_{j=1}^m \kappa(g^{-1}x_j)\, v(x_j)\, q_j.
\end{align}
Here, the group action is applied analytically to $\kappa$, while the integral is discretized.
\end{definition}

For a discrete set of output locations $\{g_i\}$, this becomes a matrix-vector product:
\begin{equation}\label{eq:disco_convolution}
\sum_{j=1}^m \kappa(g_i^{-1}x_j)\, v(x_j)\, q_j = \sum_{j=1}^m K_{ij}\, v(x_j)\, q_j,
\end{equation}
with $K_{ij}=\kappa(g_i^{-1}x_j)$. When $\kappa$ is compactly supported, $K_{ij}$ is sparse, with sparsity determined by the grid resolution and kernel support. A learnable filter is obtained by parameterizing $\kappa$ as a linear combination of a chosen set of basis functions.

For comparison, consider a standard convolutional layer with stride 1, $n$ input channels, a single output channel, and kernel $K=(K_i)_{i=1}^S\subset \mathbb{R}^n$ (with odd size $S$). On a regular grid $D_h=\{x_j\}_{j=1}^m\subset \mathbb{R}$ with spacing $h$, the output at $y\in D_h$ is given by
\begin{align}
\mathrm{Conv}_K[v](y) = \sum_{i=1}^S K_i \cdot v\Bigl(y + z_i\Bigr),
\end{align}
with $z_i = h\Bigl(i-1 - \frac{S-1}{2}\Bigr)$,
and zero-padding.%
We see that $h\to 0$, $ \lim_{h \to 0} \, \operatorname{Conv}_K[v](y)= \bar{K} \cdot v(y) \quad \text{with} \quad \bar{K} = \sum_{i=1}^S K_{i},$ this means the convolutional layer is converging to a {pointwise linear operator} as the receptive field with respect to the underlying domain $D$ is shrinking to a point. DISCO, however, does not converge to the pointwise operator.

\subsection{Kernel Basis}%

In our DISCO framework, the kernel $\kappa$ is parameterized using a basis for $L^2(\mathbb{D})$. The piecewise-linear, Zernike, and Morlet kernels are all parameterized by bases for $L^2(\mathbb{D})$. We show a specific case, using the (complex) Zernike polynomials, defined by
\begin{align}
    V_n^l(x,y) = R_n^l(\rho)e^{il\varphi}, \quad x=\rho\cos\varphi,\; y=\rho\sin\varphi,
\end{align}
where $n$ is the total degree, $|l|\le n$, and $n-|l|$ is even. The radial polynomials $R_n^l(\rho)$ satisfy
\begin{align}
\int_0^1 R_n^l(\rho)R_m^l(\rho)\,\rho\,d\rho = c_n^l\,\delta_{n,m},
\end{align}
for some nonzero constant $c_n^l$. There are exactly
$\frac{(n+1)(n+2)}{2}$
linearly independent polynomials of degree $\leq n$, so every monomial $x^i y^j$ is a finite linear combination of Zernike polynomials. By the Weierstrass theorem, they form a complete basis for $L^2(\mathbb{D})$. (More details are in Appendix VII of \cite{born2013principles}.)

\end{document}